\documentclass[prb,twocolumn,showpacs,preprintnumbers,amsmath,amssymb]{revtex4-1}

\usepackage{graphicx}
\usepackage{epstopdf}
\usepackage{dcolumn}
\usepackage{bm}

\begin{document}


\title{Single crystal growth by self-flux method of the mixed valence gold halides Cs$_2$[Au$^{I}$X$_2$][Au$^{III}$X$_4$] (X=Br,I)}

\author{Scott C. Riggs, M. C. Shapiro,  F. Corredor, T. H. Geballe, I. R. Fisher} 
\affiliation{Department of Applied Physics and Geballe Laboratory for Advanced Materials, Stanford University, CA 94305.}
\author{Gregory T. McCandless, Julia Y. Chan}
\affiliation{Department of Chemistry, Louisiana State University, Baton Rouge, LA 70803}

\date{\today}

\begin{abstract}
High quality single crystals of Cs$_2$Au$_2$X$_6$ (X=Br,I) were grown using a ternary self-flux method.  Structural refinements based on single crystal X-ray diffraction measurements show that both materials have a distorted perovskite structure belonging to the I4/mmm space group with full site occupancy. Transport measurements reveal a large bandgap of 550 $\pm$ 100 meV for Cs$_2$Au$_2$I$_6$ and 520 $\pm$ 80 meV for Cs$_2$Au$_2$Br$_6$. Initial attempts at chemical substitution are described. 
\end{abstract}
\pacs{???, ???, ???}
\maketitle

\section{introduction}

The title compounds, Cs$_2$Au$_2$X$_6$ (X=Br,I) adopt a distorted perovskite structure. Although simple charge counting arguments might indicate a single gold valence Au$^{II}$, and hence a metallic state, the materials are in fact found to be hard insulators, comprising two distinct Au sites with different formal valences Au$^I$ and Au$^{III}$ \cite{tanino86, kojima91, kojima97, ikeda04, liu04}. However, hydrostatic pressure can induce a coupled first order structural and valence transition at $\approx$9 GPa for X=Br and $\approx$5.5 GPa for X=I, spurring considerable recent interest in the associated changes in the electronic behavior \cite{kojima97, kojima99, kojima00}. 

The mixed valence state in Cs$_2$Au$_2$X$_6$ at ambient pressure is reminiscent of  Ba$_2$Bi$_2$O$_6$, which also adopts a distorted perovskite structure in which two distinct Bi sites have been associated with formal valences Bi$^{III}$ and Bi$^V$ \cite{cox79}. Significantly, in the case of Ba$_2$Bi$_2$O$_6$, suppression of the charge density wave (CDW) state via hole doping yields superconducting ground states for Ba$_{2-x}$K$_x$Bi$_2$O$_6$ \citep{cava88,pei90} and Ba$_2$Bi$_{2-x}$Pb$_x$O$_6$ \cite{sleight75} with moderately high maximum $T_c$ values of 30 and 12 K, respectively. The origin of the pairing interaction in these compounds is unclear and is still a matter of ongoing research. One possibility that has been suggested is that the correlations that result in CDW formation for the parent compound are still present for the doped materials, and that the associated valence fluctuations in the metallic regime result in a local pairing interaction \cite{varma88}. Some evidence such a scenario might be appropriate can be found in the doped superconducting semiconductor Pb$_{1-x}$Tl$_x$Te, for which Kondo-like behavior in the absence of magnetic impurities has been attributed to Tl valence fluctuations \cite{fisher05, fisher2009, dzero05}. The physical origin of the charge disproportionation in these materials is likely somewhat different, though. In the case of Ba$_2$Bi$_2$O$_6$ and Tl-doped PbTe, the two valence states are associated with filled and empty $6s$ shells, whereas for Cs$_2$Au$_2$X$_6$ Jahn-Teller distortion of the coordinating halogen ligands stabilizes $5d^8$ and $5d^{10}$ electron configurations. Nevertheless, the fact that charge disproportion plays an important role for both materials motivates a concerted experimental effort aimed at suppressing the mixed valence CDW state in Cs$_2$Au$_2$X$_6$ via chemical substitution. Controllable chemical substitution requires a synthesis technique that readily allows additional elements to be introduced in a controlled fashion. 

Previously, single crystals of Cs$_2$Au$_2$X$_6$ (X = Br and I) have been grown via a diffusion method, using either acetonitrile or hydriodic acid as a solvent respectively \cite{rau73, kojima00}. This technique suffers the drawback of not being easy to introduce additional elements to the solution, motivating development of alternative crystal growth techniques. Here we show how large, high quality single crystals of Cs$_2$Au$_2$X$_6$ can be grown directly from a ternary melt via a slow-cooling, self flux technique. We characterize the samples by single crystal x-ray diffraction, electron microprobe analysis, and electrical transport measurements. We also describe results of initial attempts at chemical substitution.

\section{Synthesis}

\begin{figure}[ht]
\centering
\includegraphics[bb=0 0 800 150,scale=1]{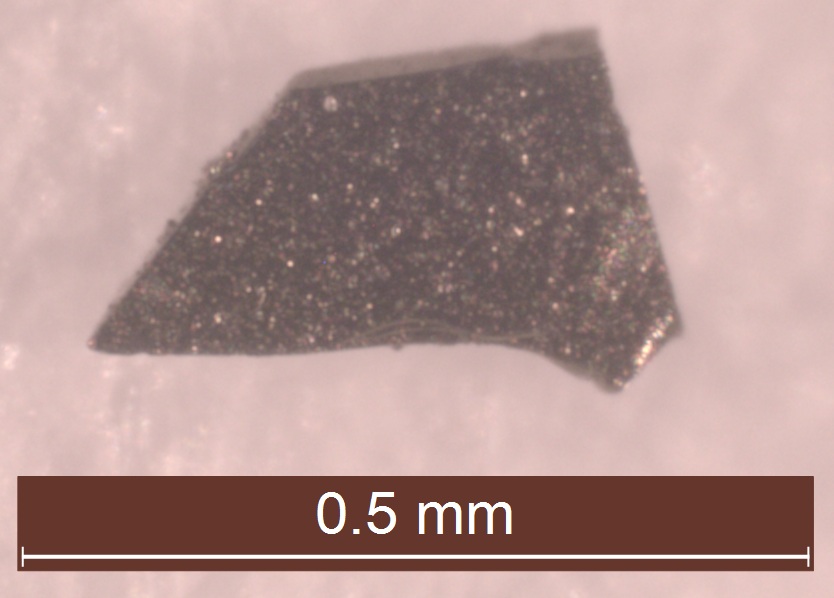}
\caption{Photograph of a representative single crystal of Cs$_2$Au$_2$I$_6$ grown via a self flux. The crystals form as thin plates, with the c-axis perpendicular to the plane of the plates, and are gold in color.}
\label{fig:CsAuI}
\end{figure}

The largest crystals of Cs$_2$Au$_2$I$_6$ were obtained by slow cooling mixtures of CsI (Alfa, 99.999$\%$), Au (Alfa, .99$\%$), and I$_2$ (Alfa, 99.99$\%$) in molar ratios of 1.3:1:1 respectively.  The reagents were weighed and placed in a 2 ml alumina crucible, which was then sealed in a quartz tube after being flushed with argon and evacuated. The quantity of reagents was calculated such that the pressure in the quartz tube would not exceed 3 atm at the maximum temperature. For Cs$_2$Au$_2$Br$_6$, optimal results were obtained from a similar molar ratio of reagents (CsBr: Alfa, 99.999$\%$), but liquid bromine (Alfa, 99.998$\%$) was added to the crucible inside the quartz tube using a micro-pipette $\it{after}$ the tube was partially necked using a hydrogen torch to minimize loss of bromine. The end of the quartz tube holding the crucible was then cooled by immersion in liquid nitrogen to avoid evaporation of Br during the evacuation and subsequent sealing of the ampoule. The sealed quartz tubes were placed in a furnace and heated from room temperature to 630$^o$C for Cs$_2$Au$_2$Br$_6$ and 550$^o$C for Cs$_2$Au$_2$I$_6$ over the course of 42 hours. After dwelling for 10 hours, the furnace was allowed to cool back to room temperature at a rate of 5 C/hr. Crystal growth does not appear to occur via chemical vapor transport, but rather via precipitation from the molten solution. Care must be taken when opening the quartz tube after the growth procedure since small quantities of unreacted Cs can sometimes react violently when exposed to oxygen. Crystals with dimensions up to approximately $600 \mu m$ x $300 \mu m$ x $50 \mu m$ can be readily extracted from the crucible. The crystals have a platelike morphology, with the c-axis perpendicular to the plane of the plate and are gold in color (Figure \ref{fig:CsAuI}). The material reacts with moisture, and must be stored in a dessicator.

The crystal growth method described above readily enables incorporation of a wide variety of additional elements to the melt. Initial substitution experiments included Zn, Ga, Se, Ag, Cd, In, Te, Pt, Pb, Bi and Ho, for a variety of concentrations. In all cases, the additional elements appeared to dissolve in the melt. And in nearly all cases, slow cooling of the quaternary melt still resulted in crystal growth of Cs$_2$Au$_2$X$_6$, though often with slightly reduced size. However, microprobe analysis revealed that none of these elements were incorporated into the Cs$_2$Au$_2$X$_6$ crystal lattice to any appreciable extent. Furthermore, electrical resistivity measurements for crystals taken from these batches always revealed insulating behavior with a similar gap to that of the parent compounds (see section IV below). Hence, despite considerable effort, we have thus far been unable to make Cs$_2$Au$_2$X$_6$ metallic by chemical substitution. The list of elements that we have been able to try thus far is certainly not exhaustive, and several other impurities and other strategies (including high pressure synthesis) might be more favorable. 

The remainder of the paper is dedicated to the characterization of the undoped parent compounds grown from ternary melts as described above.

\section{Crystal Structure}
Initial powder x-ray diffraction measurements indicated a similar structure to previous measurements of Cs$_2$Au$_2$X$_6$ \cite{kojima97}. For a full structure determination, crystallographic data was collected using a Nonius KappaCCD X-ray diffractometer equipped with a Mo K$\alpha$ radiation source ($\lambda$ = 0.71073 angstrom) and a graphite monochromator. For Cs$_2$Au$_2$I$_6$, a crystal with dimensions $\approx$ 0.05mm x 0.35mm x 0.45mm was mounted on a thin glass fiber using epoxy as an adhesive and diffraction data was collected out to 2$\theta$ of 80 degrees. The data collection strategy (a combination of $\phi$ and $\omega$ scans) was developed based on preliminary frames collected for unit cell determination and the criteria of a desired dataset completeness of 99.995$\%$ and redundancy of 3.  Unit cell parameters and systematic absences suggested that the structure is body-centered tetragonal.  Absorption was corrected using the multi-scan correction \cite{minor97}.  After determining the space group with the maXus software package, a preliminary model was generated using SIR97 \cite{altomare99}. Final least-squares refinement of the model and extinction correction was carried out in SHELXL-97 and missing symmetry was evaluated with PLATON \cite{acta08, spek03}. The best fitting space group for Cs$_2$Au$^{I}$Au$^{III}$I$_6$ is I4/mmm (No. 139) and our model agrees with previous structure report for this compound \cite{kojima97}. 

A similar structure determination was also performed for Cs$_2$Au$^I$Au$^{III}$Br$_6$ using x-ray diffraction data collected with a smaller single crystal, $\approx$ 0.02mm x 0.1mm x 0.2mm). The best fitting space group for Cs$_2$Au$^I$Au$^{III}$Br$_6$ is also I4/mmm (No. 139) and coincides with previous structure reports for this compound \cite{kojima05}.

Details of the structural refinement for both compounds are listed in Tables I-IV. For both cases, the x-ray refinement indicates full site occupancy. For Cs$_2$Au$_2$I$_6$, the cation stoichiometry was also checked by electron microprobe analysis, using a JEOL JXA-733A  with elemental Cs and Au standards. Normalized to the Cs content, the Au content was found to be 1.004$\pm$0.032, confirming the full site occupancy determined by x-ray diffraction. Uncertainties represent the standard deviation between multiple measurements on the same crystal performed at different crystallographic locations. The iodine concentration could not be accurately determined due to partial decomposition in the electron beam, presumably due to heating effects. 

As illustrated in Fig.2, the Cs$_2$Au$_2$X$_6$ structure comprises a distorted perovskite structure in which linearly coordinated AuX$_2$ and square-planar coordinated AuX$_4$ complexes alternate through the crystal lattice. Interatomic distances are listed in Table IV. As previously suggested, the coordination is typical of Au$^I$ and Au$^{III}$ formal valences respectively \cite{kojima00, kojima05}. The ratios of short-to-long Au-X bonds $\frac{Au1-X1}{Au1-X2}$ and  $\frac{Au2-X2}{Au2-X1}$ is slightly closer to unity for X = I (0.7631 and 0.8017 respectively) than for Br (0.7464 and 0.7883). This difference can also be appreciated by inspection of the internal parameters describing the positions of the halogen ions X1 and X2 in the unit cell (Table III), which are closer to the fully symmetric positions (0.25, 0.25, 0) and (0.5, 0.5, 0.25) respectively for X = I than for X = Br. This difference is likely directly related to the reduced value of the critical pressure associated with the coupled valence and structural transition found for Cs$_2$Au$_2$I$_6$ ($\approx$5.5 GPa) relative to Cs$_2$Au$_2$Br$_6$ ($\approx$9 GPa), but the physical origin of the difference is unclear.

\begin{figure}[ht]
\centering
\includegraphics[bb=0 0 800 480,scale=0.3]{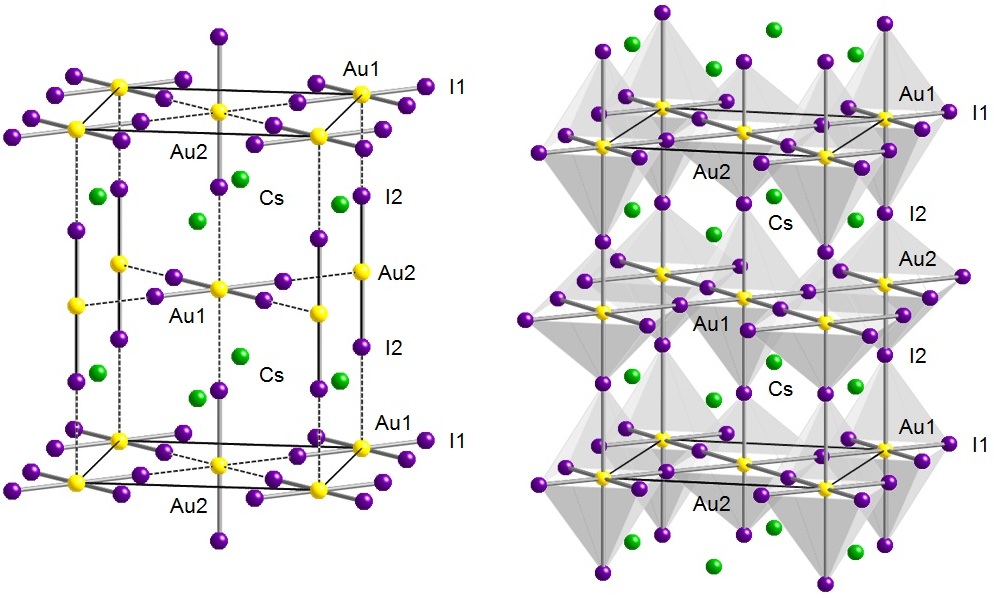}
\caption{Crystal structure of the distorted pervoskite Cs$_2$Au$^{I}$Au$^{III}$I$_6$. Cs and Au sites are labelled; X sites are shown by purple spheres. The left hand panel shows the square planar Au$^{III}$I$_4$ and linear chain Au$^{I}$I$_2$ complexes associated with the Au1 and Au2 sites respectively.   The right hand panel emphasizes the distorted octahedral coordination of the two sites (compressed octahedral coordination for the (Au$^{I}$I$_2$)$^-$ chains versus elongated octahedral coordination for the (Au$^{III}$I$_4$)$^-$ square planar complexes). The tetragonal c-axis is vertical.}
\label{fig:CsAuIstruct}
\end{figure}

\begin{table}[ht]
\caption{Crystallographic Parameters} 
\centering  
\begin{tabular}{l*{2}{l}r}
\hline
\hline
\emph{Crystal data}   & &     \\
Formula & Cs$_2$Au$_2$I$_6$ & Cs$_2$Au$_2$Br$_6$ \\
M$_r$(g mol$^{-1}$) & 1421.15 & 1139.21 \\
Crystal system & Tetragonal &  Tetragonal\\
Space group & I4/mmm & I4/mmm \\
a($\AA$) & 8.2847(10) & 7.7869(10)\\
c($\AA$) & 12.0845(15) & 11.3808(15)\\
V($\AA^3$) & 829.43(18) & 690.08(15) \\
Z & 2 & 2\\
F(000) & 1172 & 956\\
D$_X$(g cm$^{-3}$) & 5.690 & 5.483\\
2$\theta$ range($^{\circ}$) & 5.2-55.0 & 2.0-80.6\\
$\mu$ (mm$^{-1}$) & 33.12 & 43.75\\
Temperature (K) & 298(2) & 298(2)\\
Mosaicity ($^{\circ}$) & 0.494(1) & 0.544(2)\\
\\
\emph{Data collection} \\
Measured reflections & 948 & 2183\\
Independent reflections & 306 & 669 \\
Reflections with I$>$2$\sigma$(\emph{I}) & 301 & 505\\
$^a$R$_{int}$ & 0.026 & 0.036\\
\emph{h} & -10 $\rightarrow$ 10 & -14 $\rightarrow$ 14\\
\emph{k} & -7 $\rightarrow$ 7 & -9 $\rightarrow$ 10\\
\emph{l} & -15 $\rightarrow$ 15 & -20 $\rightarrow$ 20\\
\\
\emph{Refinement} \\
Reflections & 306 & 669\\
Parameters & 15 & 15\\
$^b$R$_1$[F$^2>$2$\sigma$(F$^2$)] & 0.048 & 0.028\\
$^c \emph{w}$R$_2$(F$^2$) & 0.130 & 0.075\\
$^d$S & 1.23 & 1.06\\
$\Delta \rho_{max}$(e$\AA^{-3}$) & 2.90 & 1.53\\
$\Delta \rho_{min}$(e$\AA^{-3}$) & -5.39 & -2.97\\
Extinction coefficient & 0.0034(5) & 0.00157(15)\\
\hline
\hline
\multicolumn{3}{l}{$^a$R$_{int}$ = [$\Sigma$ $|${F$_o^2$ - F$_o^2$(mean)}$|$ / $\Sigma$ F$_o^2$]}  \\
\multicolumn{3}{l}{$^b$R$_1$ = $||$F$_o|$-$||$F$_c||$ / $\Sigma|$F$_o|$} \\
\multicolumn{3}{l}{$^c \emph{w}$R$_2$={$\{\Sigma$[$\emph{w}$(F$_o^2$-F$_c^2$)$^2$]/$\Sigma$[$\emph{w}$(F$_o^2$)$^2$]}$\}^\frac{1}{2}$, } \\
\multicolumn{3}{l}{$\emph{w}$= 1/[ $\sigma^2$(F$_o^2$) + (0.0995P)$^2$ + 0.8069P ] for Cs$_2$Au$_2$I$_6$} \\
\multicolumn{3}{l}{$\emph{w}$= 1/[ $\sigma^2$(F$_o^2$) + (0.0356P)$^2$ + 2.4804P ] for Cs$_2$Au$_2$Br$_6$} \\
\multicolumn{3}{l}{$^d$S={$\{\Sigma$[$\emph{w}$(F$_o^2$ - F$_c^2$)$^2$ ]/(n-p)}$\}^\frac{1}{2}$} \\
\end{tabular}
\label{table:tablestuff} 
\end{table}

\begin{table}
\caption{Selected Interatomic Distances ($\AA$)} 
\centering  
\begin{tabular}{l*{2}{l}r}
\hline
\hline
Cs-I1 (x8) & 4.2175(4) \\
Cs-I2 (x4) & 4.1662(5) \\
Au1-I1 (x4) & 2.6454(13) \\
Au1-I2 (x2) & 3.4666(14) \\
Au2-I1 (x4)	& 3.2128(13) \\
Au2-I2 (x2)	& 2.5757(14) \\
\\
\multicolumn{2}{l}{Au1 $\rightarrow$ Au2 (along ab plane) 5.8582(7)} \\
\multicolumn{2}{l}{Au1 $\rightarrow$ Au2 (along c axis) 6.0423(8)} \\
\hline
Cs-Br1(x8) & 3.9709(4) \\
Cs-Br2(x4) & 3.9175(5) \\
Au1-Br1(x4) & 2.4471(9) \\
Au1-Br2(x2) & 3.2787(13) \\
Au2-Br1(x4) & 3.0591(10) \\
Au2-Br2(x2)& 2.4117(13) \\

\multicolumn{2}{l}{Au1 $\rightarrow$ Au2 (along ab plane) 5.5062(5)} \\
\multicolumn{2}{l}{Au1 $\rightarrow$ Au2 (along c axis) 5.6904(8)} \\

\hline
\hline

\end{tabular}
\label{table:atomicdis} 
\end{table}

\begin{table*}
\caption{Atomic Positions and Equivalent Isotropic Displacement Parameters} 
\centering  
\begin{tabular}{l*{7}{l}r}
\hline
\hline
Atom  & Wyckoff position & x & y & z & $^a$Occ. & $^b$U$_{eq}$($\AA^2$)     \\
\hline
Cs & 4$\emph{d}$ & 0 & $\frac{1}{2}$ & $\frac{1}{4}$ & 1 & 0.0543(8) \\
Au1(+3) & 2$\emph{a}$ & 0 & 0 & 0 & 1 & 0.0226(5) \\
Au2(+1) & 2$\emph{b}$ & $\frac{1}{2}$ & $\frac{1}{2}$ & 0 & 1 & 0.0269(5) \\
I1 & 8$\emph{h}$ & 0.22578(11) & 0.22578(11) & 0 & 1 & 0.0423(6)\\
I2 & 4$\emph{e}$ & $\frac{1}{2}$ & $\frac{1}{2}$ & $0.21314(11)$ & 1 & 0.0381(6) \\
\hline
Cs & 4$\emph{d}$ & 0 & $\frac{1}{2}$ & $\frac{1}{4}$ & 1 & 0.0466(2) \\
Au1(+3) & 2$\emph{a}$ & 0 & 0 & 0 & 1 & 0.02299(13) \\
Au2(+1) & 2$\emph{b}$ & $\frac{1}{2}$ & $\frac{1}{2}$ & 0 & 1 & 0.02541(14) \\
Br1 & 8$\emph{h}$ & 0.22221(8) & 0.22221(8) & 0 & 1 & 0.0402(2) \\
Br2 & 4$\emph{e}$ & 1/2 & 1/2 & 0.21191(11) & 1 & 0.0385(3) \\

\hline
\hline
\multicolumn{7}{l}{$^a$Occupancy of atoms} \\
\multicolumn{7}{l}{$^b$U$_{eq}$ is defined as one-third of the trace of the orthogonalized U$^{ij}$ tensor}
\end{tabular}
\label{table:atomicpos} 
\end{table*}

\begin{table*}
\caption{Anisotropic Atomic Displacement Parameters ($\AA^2$)} 
\centering  
\begin{tabular}{l*{7}{l}r}
\hline
\hline
Atom  & U$^{11}$ & U$^{22}$ & U$^{33}$ & U$^{12}$ & U$^{13}$ & U$^{23}$ \\
\hline
Cs & 0.0537(11) & 0.0537(11) & 0.0553(15) & 0.000 & 0.000 & 0.000\\
Au1 & 0.0232(6) & 0.0232(6) & 0.0215(8) & 0.000 & 0.000 & 0.000\\
Au2 & 0.0318(6) & 0.0318(6) & 0.0169(7) & 0.000 & 0.000 & 0.000\\
I1 & 0.0376(7) & 0.0376(7) & 0.0518(9) & -0.0140(5) & 0.000 & 0.000\\
I2 & 0.0485(8) & 0.0485(8) & 0.0173(9) & 0.000 & 0.000 & 0.000\\
\hline
Cs & 0.0461(3) & 0.0461(3) & 0.0475(5) & 0.000 & 0.000 & 0.000 \\
Au1 & 0.02288(16) & 0.02288(16) & 0.0232(2) & 0.000 & 0.000 & 0.000 \\
Au2 & 0.02714(17) & 0.02714(17) & 0.0219(2) & 0.000 & 0.000 & 0.000 \\
Br1 & 0.0358(3) & 0.0358(3) & 0.0490(5) & -0.0118(3) & 0.000 & 0.000 \\
Br2 & 0.0468(4) & 0.0468(4) & 0.0220(4) & 0.000 & 0.000 & 0.000 \\

\hline
\hline
\end{tabular}
\label{table:atomicbla} 
\end{table*}

\section{Electrical Resistivity}

Transport measurements were made on single crystals of Cs$_2$Au$_2$X$_6$ using a Keithley electrometer.  Samples were mounted in a 2-wire configuration for measurement of the in-plane resistivity. Contacts were made with Au wires using Dupont 4929 Ag-epoxy and allowed to set at room temperature.  Samples were attached to a glass slide and wired to an Oxford Instruments optiflow cryostat probe with two separated coaxial cables.  The inner cores of the coaxial cables were used for the voltage measurement. Measurements were made over a temperature range of roughly 300K to 200K with the sample in He exchange gas.  Below $\approx$200K the resistivity became too large and was no longer measurable using the current experimental set-up.  

\begin{figure}[ht]
\centering
\includegraphics[bb=0 0 640 780,scale=0.75]{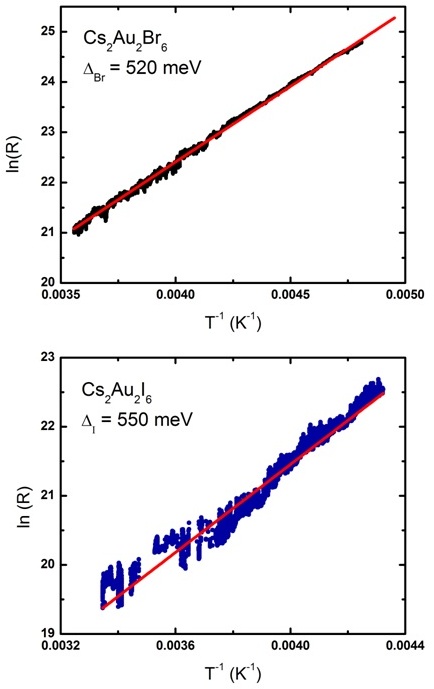}
\caption{Arhenius plots showing activated behavior of the in-plane resistance of representative single crystals of (a) Cs$_2$Au$_2$Br$_6$ and (b) Cs$_2$Au$_2$I$_6$. Red lines show linear fits, yielding gaps of $\Delta_{I}$ is 550 $\pm$ 100 meV and $\Delta_{Br}$ is 520 $\pm$ 80 meV.}
\label{fig:gapfit}
\end{figure}

The temperature dependence of the resistivity of both compounds follows an activated behavior, which can be readily appreciated by inspection of figure \ref{fig:gapfit} showing the natural log of the resistance as a function of inverse temperature. Data plotted in this manner were fit to a straight line and the gap $\Delta$ determined assuming typical semiconductor activated behavior:

\begin{equation}
R=R_0 e^{\frac{\Delta}{2k_{B}T}} 
\label{eq:semiconduct}
\end{equation}

Resulting values of the bandgap are, $\Delta_{I}$ = 550 $\pm$ 100 meV and $\Delta_{Br}$ = 520 $\pm$ 80 meV for Cs$_2$Au$_2$I$_6$ and Cs$_2$Au$_2$Br$_6$, respectively, where the relatively large uncertainty reflects a combination of electrical contact to the material and sample-to-sample variation.

The gold-halide materials are very good insulators, with a typical room temperature resistivity of $\approx$700 m$\Omega$ cm, corresponding to resistances of 500 M$\Omega$ for typical sample dimensions.  The measured transport gap (550 meV for X=I) is considerably less than estimates of the minimum intervalence charge transfer (IVCT) excitation determined from optical reflectivity measurements ($\approx$ 806 meV for the Au$^{I}$(5d$_{z^2}$) to Au$^{III}$(5d$_{x^2-y^2}$) transition for Cs$_2$Au$_2$I$_6$ observed for electric fields oriented perpendicular to the crystallographic c-axis) determined from optical reflectivity measurements \cite{kojima99}. 
  
  The reason for this difference is not immediately obvious. Given the very large value of the intrinsic resistance of the samples (rising up to 25 G$\Omega$ at $\approx$200K for typical samples), the discrepancy might be due to excess halogen leaching out of the crystals, or water settling on the sample surface, affecting current paths and lowering the effective gap. Unfortunately, the surfaces could not be easily cleaned with usual solvents, including water, methanol, acetone, and isopropyl alcohol, due to adverse reaction. Equally, a difference in the apparent gap probed by transport and reflectivity measurements can arise from a number of mechanisms including an indirect bandgap, pinning of the chemical potential by localized impurity states, or perhaps if the charge transport mechanism does not involve single carrier hopping. Perhaps coincidentally, the closely related CDW compound Ba$_2$Bi$_2$O$_6$ also exhibits a significant difference in the transport and optical gap (by a factor of almost ten) \citep{sugai87}, although the origin of this effect is uncertain. \

\section{Conclusions}
In conclusion, we have described an alternative method to grow large, high quality, single crystals of the mixed valence compounds Cs$_2$Au$^{I}$Au$^{III}$X$_6$, (X=I,Br) by means of a ternary self-flux.  Single crystals x-ray diffraction measurements confirmed the previously determined crystal structure of both compounds\cite{kojima97, kojima05}. Transport measurements reveal a band gap of 550 $\pm$ 100 meV for Cs$_2$Au$_2$I$_6$ and 520 $\pm$ 80 meV for Cs$_2$Au$_2$Br$_6$. Initial attempts to chemically substitute with a variety of dopants, were unsuccessful. 

\section{Acknowledgments}
Scott Riggs would like to thank R. Jones for microprobe analysis. Work at Stanford University was supported by the AFOSR grant FA9550-09-1-0583. Work at LSU was supported by NSF grant DMR 1063735.

\bibliography{CsAuI}



\end{document}